\documentclass[useAMS,usenatbib]{mn2e}
\usepackage{graphicx}

\title[Follow-up photometry of TrES-3]{Photometric follow-up of the transiting planetary system TrES-3:
transit timing variation and long-term stability of the system
\thanks{Partly based on observations made at the Centro Astron\'omico
Hispano Alem\'an (CAHA), operated jointly by the Max-Planck Institut f\"ur Astronomie 
and the Instituto de Astrof\'{\i}sica de Andaluc\'{\i}a (CSIC).
}}
\author[M. Va\v{n}ko et al.]
  {M.~Va\v{n}ko,$^1$\thanks{vanko@ta3.sk}
G.~Maciejewski$^2$, M.~Jakub\'ik$^1$, T.~Krej\v{c}ov\'a$^3$, J.~Budaj$^1$,
  \newauthor 
  T. Pribulla$^1$, J.~Ohlert$^{4,5}$,
  St.~Raetz$^6$, \v{S}.~Parimucha$^7$, L.~Bukowiecki$^2$\\
  $^1$Astronomical Institute, Slovak Academy of Sciences,
      059 60 Tatransk\'a Lomnica, Slovakia\\
  $^2$Toru\'n Centre for Astronomy, Nicolaus Copernicus University, Gagarina 11,
      87100, Toru\'n, Poland\\
  $^3$Masaryk University, Department of Theoretical Physics and
      Astrophysics, 602 00 Brno, The Czech Republic\\
  $^4$University of Applied Sciences, Wilhelm-Leuschner-Strasse 13, 61169
Friedberg,
      Germany\\
  $^5$Michael Adrian Observatory, Astronomie Stiftung Trebur, Fichtenstrasse 7,
65468 Trebur, 
      Germany\\
  $^6$Astrophysikalisches Institut und Universit\"ats-Sternwarte,
      Schillerg\"a{\ss}chen 2-3, 07745 Jena, Germany\\
  $^7$Institute of Physics, Faculty of Natural Sciences, \v{S}af\'arik
      University, Jesenn\'a 5, 04001, Ko\v{s}ice, Slovakia
}

\begin{document}

\date{Accepted 2013 December 15. Received 2013 December 6; in original form
2012 October 11}

\pagerange{\pageref{firstpage}--\pageref{lastpage}} \pubyear{2012}

\maketitle

\label{firstpage}

\begin{abstract}
We present new observations of the transiting system TrES-3
obtained from 2009 to 2011 at several observatories.
The orbital parameters of the system were re- determined and a new linear
ephemeris was calculated. The best quality light curve was used for light curve analysis, and 
other datasets were used to determine mid-transit times ($T_{C}$) and study 
transit time variation (TTV). For planet parameter determination we used two independent 
codes and finally, we concluded that our parameters are in agreement with
previous studies. Based on our observations, we determined 14 mid-transit times.
Together with published $T_{C}$ we found that the timing residuals showed no significant 
deviation from the linear ephemeris. We concluded that a periodic TTV signal with
an amplitude greater than 1 minute over a 4-year time span seems to be unlikely. 
Our analysis of an upper mass limit allows us to exclude an additional Earth-mass 
planet close to inner 3:1, 2:1, and 5:3 and outer 3:5, 1:2, and 1:3 mean-motion 
resonances. 
Using the long-term integration and applying the method of maximum
eccentricity, the region from 0.015$\,$au to 0.05$\,$au was found unstable and 
the region beyond the 0.05$\,$au was found to have a chaotic behaviour 
and its depletion increases with increasing values of
the initial eccentricity as well as inclination.

\end{abstract}


\section{Introduction}

Since the first discovery of a transiting planet around HD~209458
\citep{charbo+00}, 235 transiting extrasolar systems have already been
confirmed up to December 4th, 2012.\footnote{http://exoplanet.eu}
Whilst transiting exoplanets offer unique scientific
possibilities, their study involves several complications. 
In general, it is impossible to measure the mass
and radius of a planet based on a dataset obtained    
with one observational technique. Transit light curves allow us to 
determine just the relative
size of a star and planet, the orbital inclination and the stellar 
limb-darkening coefficients. By combining this with radial-velocity measurements, 
the observations offer the opportunity to measure the precise 
stellar and planetary parameters. In order to obtain such parameters, 
some constraints are needed, and are usually provided by forcing the 
properties of the host stars to match theoretical expectations
\citep{southworth10}.
Significant uncertainties remain in the stellar mass and radius
determinations of many systems. In some cases, this is due to poorly 
determined photospheric properties (i.e. effective temperature and metallicity), 
and in other cases due to a lack of an accurate luminosity estimate
\citep{sozzetti+09}. In addition, the different methods used for 
these determinations as well as different approaches toward systematic 
errors are leading to rather inhomogeneous set of planet properties. 
Because of such inhomogenities, recently a few papers were published
where authors re-analyzed a large subset of known transiting planets,
and applied a uniform methodology
to all systems (e.g. \citealp{torres+08, southworth10, southworth12}).

In this paper we focus on the transiting system TrES-3.
The system consists of a nearby G-type dwarf and a massive
hot Jupiter with an orbital period of 1.3 days. It was discovered by
\citet{odonovan+07} and also detected by the SuperWASP
survey \citep{collier07}. Later, \citet{sozzetti+09} presented new spectroscopic
and photometric observations of the host star.
A detailed abundance analysis based on high-resolution spectra 
yields [Fe/H] = $-$0.19 $\pm$ 0.08, $T_{\rm eff}$~=~5650 $\pm$ 75~K and 
log~$g$~=~4.0~$\pm$~0.1. The spectroscopic orbital solution was improved 
with new radial velocity measurements obtained by \citet{sozzetti+09}. 
Moreover, these authors redetermined the stellar parameters 
(i.e. $M_{*}$ = 0.928$^{+0.028}_{-0.048}$
$M_\odot$ and $R_{*}$ = 0.829$^{+0.015}_{-0.022}$ $R_\odot$) and finally, 
the new values of the planetary mass and radius were determined (see Tab \ref{tab03}). 
They also studied the transit 
timing variations (TTVs) of TrES-3 and noted significant outliers from a constant
period. In the same year, \citet{gibson09} presented the follow-up transit 
photometry. It consisted of nine transits of TrES-3, taken as part of
transit timing program using the RISE instrument on the Liverpool Telescope. 
These transits, together with eight transit times published before \citep{sozzetti+09}, were 
used to place upper mass limit as a function of the period ratio of a
potential perturbing planet and transiting planet. It was shown that 
timing residuals are sufficiently sensitive to probe sub-Earth mass planet in both interior
and exterior 2:1 resonances, assuming that the additional planet is in an
initially circular orbit. \citet{christiansen+11} has observed TrES-3 as a
part of the NASA {\it EPOXI} Mission of Opportunity. They detected 
a long-term variability in the TrES-3 light curve, which may be due to 
star spots. They also confirmed that the planetary 
atmosphere does not have a temperature inversion. 
Later, \citet{turner+13} observed nine primary transits of the hot Jupiter 
TrES-3b in several optical and near-UV photometric bands from June 2009 to April
2012 in an attempt to detect its magnetic field. Authors determined an upper limit of
TrES-3b's magnetic field strength between 0.013 and 1.3 G using a timing difference of 138 s
derived from the Nyquist--Shannon sampling theorem. They also presented a
refinement of the physical parameters of TrES-3b, an updated
ephemeris and its first published near-UV light curve. The near-UV planetary
radius of $R_{p}$ = 1.386$^{+0.248}_{-0.144}$ $R_{J}$ was also determined. 
This value is consistent with the planet's optical radius.
Recently, \citet{kundurthy+13} observed eleven transits of TrES-3b over
a two year period in order to constrain system parameters and look for transit timing 
and depth variations. They also estimated the system parameters for 
TrES-3b and found consistency with previous estimates. Their analysis of the
transit timing data show no evidence for transit timing variations and timing 
measurements are able to rule out Super-Earth and Gas Giant companions in low order 
mean motion resonance with TrES-3b. 

The main aims of this study can be summarized in the following items:
(i) determination of the system parameters for TrES-3b (two independent
codes will be used) and comparison with previous studies 
(i.e. \citealp{odonovan+07, sozzetti+09, gibson09, colon+10, southworth10,
southworth11, lee11, christiansen+11, sada+12, turner+13, kundurthy+13}).
(ii) based on the obtained transits, we will determine the mid-transit times ($T_{C}$) 
and with following analysis of transit time variation (TTV) we will discuss
possible presence of a hypothetical additional planet (perturber). We will
try to estimate its upper-mass limit as a function of orbital periods ratio
of transiting planet and the hypothetical perturber. (iii) Finally,     
using the long-term integration and applying the method of maximum
eccentricity we will search for stability of regions inside the TrES-3b
planet in context of additional planet(s).

The remainder of this paper is organized as follows. In the Section~2, 
we describe observations and data reduction pipelines used 
to produce the light curves. Section~3 presents the methods
for analysis of transit light curves as well as discussion and comparison 
of the parameters of TrES-3 system. Section~4 and 5 are devoted to TTV and long-term 
stability of the system, respectively. Finally, in Section~6 we summarize and discuss 
our results.  

\section[]{Observations and data reduction}

We obtained our data using several telescopes with different instruments.
This allowed us to obtain many light curves
since this strategy can effectively cope with the weather problems.   
On the other hand, this approach results in rather heterogeneous data.
We used most of the data in average quality for the TTV analysis which is
not very demanding on homogeneity of the data.
Only the best quality light curve was used for the planet parameter
determination.

Observations used in this paper were carried out at the several
observatories in Slovakia (Star\'a Lesn\'a Observatory; 49\degr 09' 10"N,
20\degr 17' 28"E), Poland (Piwnice Observatory; 53\degr 05' 43"N,
18\degr 13' 46"E), Germany (Grossschwabhausen Observatory; 50\degr 55' 44"N, 
11\degr 29' 03"E; Volkssternwarte Kirchheim Observatory, 50\degr 55'44"N, 
11\degr 29' 03"E and Michael Adrian Observatory, 49\degr 55'27"N, 08\degr
24' 33"E) and Spain (Calar Alto Observatory; 37\degr 13'25"N, 
02\degr 32' 46"E). We collected 14 transit light curves obtained
between May 2009 and September 2011. The transits on May 12, 2009 and August 20,
2010 were observed simultaneously at two different observatories. The
telescope diameters of 0.5 to 2.2 m allowed us to obtain photometry with 1.2
-- 7.8 mmag precision, depending on observing conditions. 
Observations generally started $\sim$ 1 hour before the expected beginning of a 
transit and ended $\sim$ 1 hour after the event. Unfortunately, weather
conditions and schedule constraints meant that we were not able to fit this scheme in
all cases. 

All instruments are equipped with CCD cameras with the Johnson-Cousins
($UBVR_{C}I_{C}$) 
standard filter system. 
The information from individual observatories and instruments as well as the
summary of observing runs are given in Table \ref{tab01} and Table
\ref{tab02}. The standard correction procedure 
(bias, dark and flat field correction) and subsequently aperture photometry was
performed by {\tt IRAF\/}\footnote{IRAF is distributed by the National
Optical Astronomy Observatories, which are operated by the Association of Universities for
Research in Astronomy, Inc., under cooperative agreement with the National Science
Foundation.} and task {\it chphot\/} \citep {raetz+09} (GSH and VK),
{\tt C-munipack\/} package\footnote{http://c-munipack.sourceforge.net/} (G1) and 
{\tt Mira\_Pro\_7\/}\footnote{http://www.mirametrics.com/mira\_pro.htm} (MA). Data from 
remaining telescopes (P and CA) were reduced with the software pipeline
developed for the Semi--Automatic Variability Search sky survey \citep{niedzielski+03}. 
To generate an artificial comparison star, at least 20--30 per cent of stars
with the lowest light--curve scatter were selected iteratively from the field
stars brighter than 2.5--3 mag below the saturation level 
(e.g. \citealp{broeg+05, raetz+09}). To measure
instrumental magnitudes, various aperture radii were used. The aperture which
was found to produce light curve with the smallest overall scatter was applied to
generate final light curve. The linear trend in the out-of-transit
parts was also removed.

\begin{table}
\caption{Overview of the telescopes and instruments/detectors used to obtain
photometry of TrES-3. FoV is the field of view of the instrument and
N$_{tr}$ is the number of observed transits. Abbreviations of the observatories:
{\bf G1\/} -- Star\'a Lesn\'a Observatory, {\bf GSH\/} --
Gro{\ss}schwabhausen observing station of the Jena University (CTK -- Cassegrain Teleskop
Kamera; STK -- Schmidt Teleskop Kamera, see \citealp{mugi09, mugi+10}), 
{\bf MA\/} -- Michael Adrian Observatory in Trebur, {\bf VK\/} -- 
Volkssternwarte Kirchheim Observatory (RCT -- Ritchie Chr\'etien Telescope), 
{\bf P\/} -- Piwnice Observatory and {\bf CA\/} -- Calar Alto Observatory (RCF
-- Ritchie Chr\'etien Focus). \label{tab01}}
\footnotesize
\begin{center}
\begin{tabular}{lccc}
\hline
\hline
Obs. & Telescope &  Detector & N$_{tr}$      \\
            &           &  CCD size & FoV [arcmin]   \\
\hline
G1          & Newton    &  SBIG ST10-MXE            &       5            \\
            & 508/2500  &  2184 $\times$ 1472, 6.8 $\mu$m  & 20.4 $\times$
13.8\\ 
MA          & Cassegrain&  SBIG STL-6303E           &           2         \\ 
            & 1200/9600 &   3072 $\times$ 2048, 9 $\mu$m   & 10 $\times$ 7\\ 
GSH         &   CTK     &  SITe TK1024              &         1            \\
            & 250/2250  &  1024 $\times$ 1024, 24 $\mu$m   & 37.7 $\times$
37.7\\
            &   STK     &  E2V CCD42-10             &         1        \\
            & 600/1758 &  2048 $\times$ 2048, 13.5 $\mu$m & 52.8 $\times$
52.8\\              
VK          &   RCT     &  STL-6303E                &         2       \\
            & 600/1800  &  3072 $\times$ 2048, 9 $\mu$m    & 71 $\times$ 52 \\
P           & Cassegrain&  SBIG STL-1001            &         1       \\
            & 600/13500 &  1024 $\times$ 1024, 24$\mu$m    & 11.8 $\times$
11.8\\
CA          &   RCF     &  SITe CCD                        &         2       \\
            & 2200/17037&  2048 $\times$ 2048, 24$\mu$m    &  18.1 $\times$ 18.1
\\
\noalign{\smallskip}
\hline
\hline
\end{tabular}
\end{center}
\end{table}

\begin{table}
\caption{Summary of the observing runs: Obs. -- Observatory according to
Table~1, $N_{exp}$ -- number of useful exposures, $t_{exp}$ -- exposure
times. The dates are given for the beginning of nights. \label{tab02}}
\footnotesize
\begin{center}
\begin{tabular}{lcccc}
\hline
\hline
Obs. & Date &  Filter & $N_{exp}$ & $t_{exp}$ (s)   \\
\hline
G1          & 2009 May 12  &$R$    &  319  &   40   \\
            & 2009 Aug 01  &$R$    &  345  &   45   \\
            & 2010 Apr 27  &$R$    &  180  &   35   \\
            & 2010 Jun 30  &$R$    &  238  &   40   \\
            & 2010 Aug 07  &$R$    &  168  &   35   \\
MA          & 2010 July 13 &$R$    &  349  &   25   \\
            & 2010 Aug 20  &$R$    &  296  &   20   \\
GSH (CTK)   & 2009 May 25  &$I$    &  138  &   80   \\
(STK)       & 2011 Mar 22  &$R$    &  131  &   60   \\
VK          & 2009 Aug 14  &$Clear$&  103  &  120   \\
            & 2010 Aug 20  &$Luminance$&320 & 30    \\
P           & 2009 May 12  &$R$    &  120  &   55   \\
CA          & 2010 Sep 06  &$R$    &  158  &  30-35 \\
            & 2011 Sep 12  &$R$    &  128  &  45    \\  
\noalign{\smallskip}
\hline
\hline                                                                       
\end{tabular}
\end{center}
\end{table}

\begin{figure*}
  \begin{center}
   \includegraphics[width=160mm,clip=]{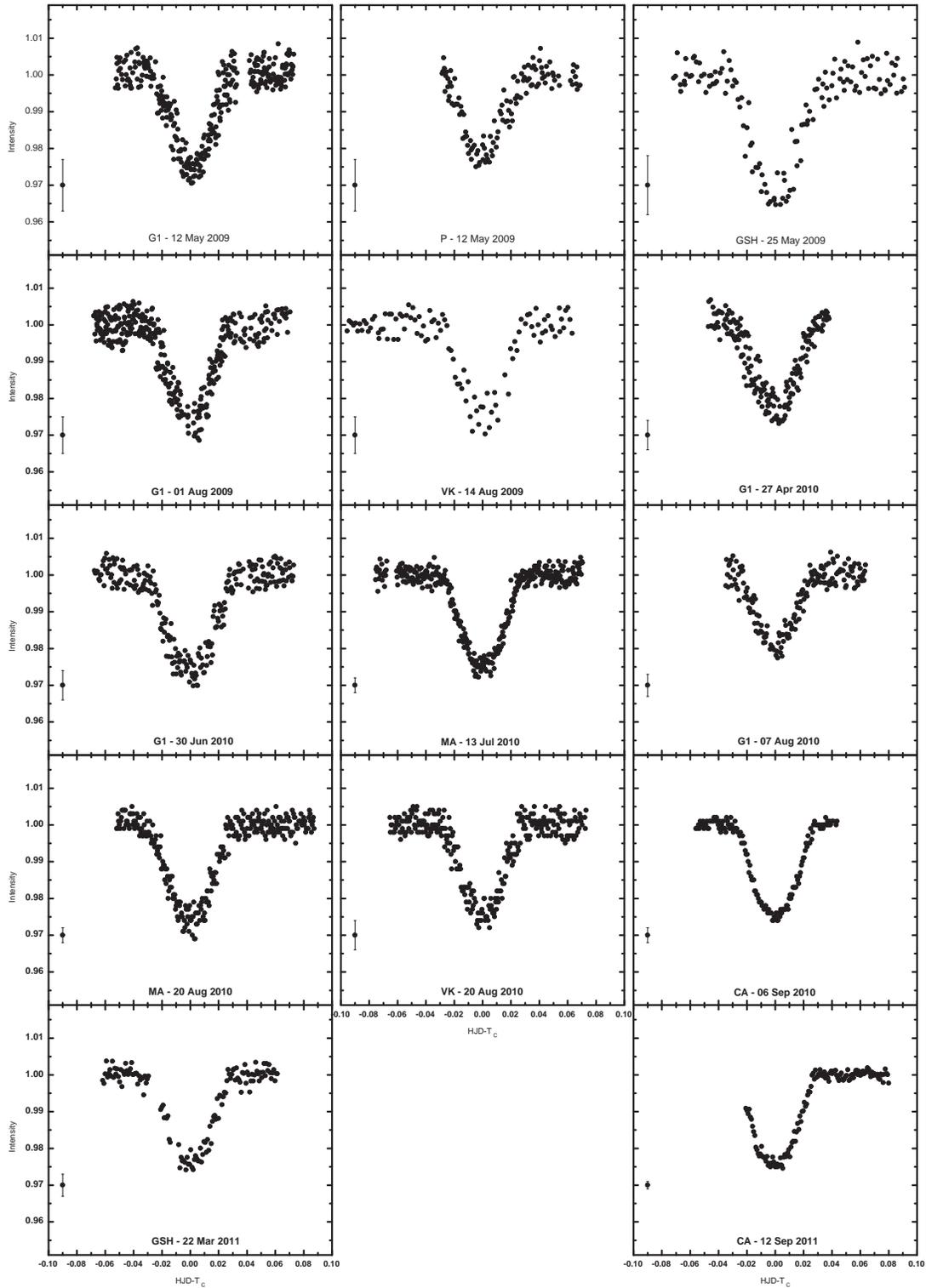}   
  \end{center}
\caption{The light curves of the system TrES-3 obtained at individual
observatories between May 2009 and September 2011. The best light curve
(obtained at Calar Alto, Sep 2010) was used to determine system parameters. 
The data from other observatories were used for $T_{C}$ determination and TTV
analysis. The typical error bar is plotted next to each light curve.\label{fig01}}
\end{figure*}

\section{Light curve analysis}

Our photometric observations of the TrES-3 system consist of data 
from different instruments and are of different
photometric quality. For the purpose of radius determination, 
we decided to analyse only a light curve with the lowest scatter and the best 
rms~=~1.2~mmag. We choose data obtained at Calar Alto observatory on September 06, 2010 (see
Figs. 1 and 2). We first refined the light-curve system parameters and subsequently determined 
the individual times of transit as described below in order to improve ephemeris. 

For the calculation of synthetic light curves we used two independent
approaches: the first one is based on the routines from \citep{mandel+02} and 
the Monte Carlo simulations (described in Section \ref{sol1}), the second one
uses {\tt JKTEBOP\/} code \citep{southworth+04} (see Section \ref{jktebop}).
\subsection{SOLUTION 1}
\label{sol1}

First we used the downhill simplex minimization procedure
(implemented in routine AMOEBA, \citealp{press+92}) to determine $4$ system parameters
$R_{\mathrm{p}}/R_{*}$ (planet to star radius ratio), $i$ (inclination), $T_{C}$ 
(mid-transit time) and $R_{*}/a$ (star radius to semi-major axis ratio). 
The model light curve itself was computed via the analytic expressions from
\citet{mandel+02}. The quadratic limb darkening law was assumed and
corresponding limb darkening coefficients $c_{1}$, $c_{2}$ were linearly interpolated from
\citet{claret00} assuming the stellar parameters from \citet{sozzetti+09}: 
$T_{\mathrm{eff}}$~=~5\,650~K, $\log$~(g)~=~4.4 and [Fe/H]~=~$-$0.19. As a goodness of the fit
estimator we used the $\chi ^2$ function:
\begin{equation}
\chi ^{2}=\sum _{i=1}^{N} \left(\frac{m_{i}-d_{i}}{\sigma_{i}}\right)^2,
\end{equation}
where $m_{i}$ is the model value and $d_{i}$ is the measured value of 
the flux, $\sigma_{i}$ is the uncertainty of the $i^{\mathrm{th}}$
measurement and the sum is taken over all measurements. 
The orbital period and the limb darkening coefficients were fixed
through the minimization procedure. 
The transit duration $T_{\mathrm{D}}$ was determined assuming the semi-major axis
$a$ = 0.02282$^{+0.00023}_{-0.00040}\,$au \citep{sozzetti+09}. The parameters and
the corresponding light curve for which we found the minimum value 
of $\chi^2$ function, are in Table 3 and Figure 2, respectively ({\it SOLUTION 1}). 

\begin{figure}
  \begin{center}
   \includegraphics[width=85mm,clip=4]{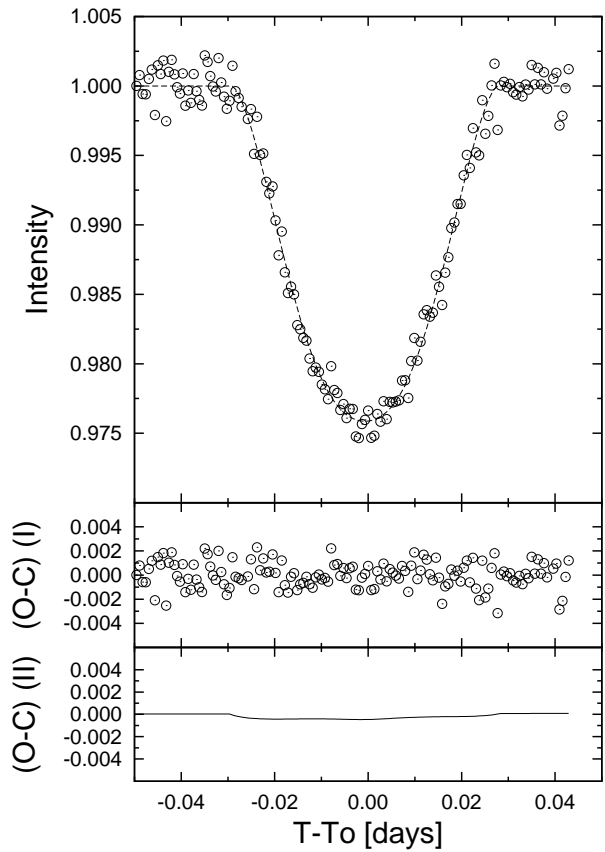}
  \end{center}
\caption{Top: The light curve obtained at Calar Alto on September 06, 2010, 
and the best fit corresponding by SOLUTION 1. Middle: Residuals from the best fit 
mentioned above. Bottom: Difference from the fits corresponding by SOLUTION 1 and 
SOLUTION 2. \label{fig02}}
\end{figure}

\begin{figure}
\includegraphics[width=80mm,clip=]{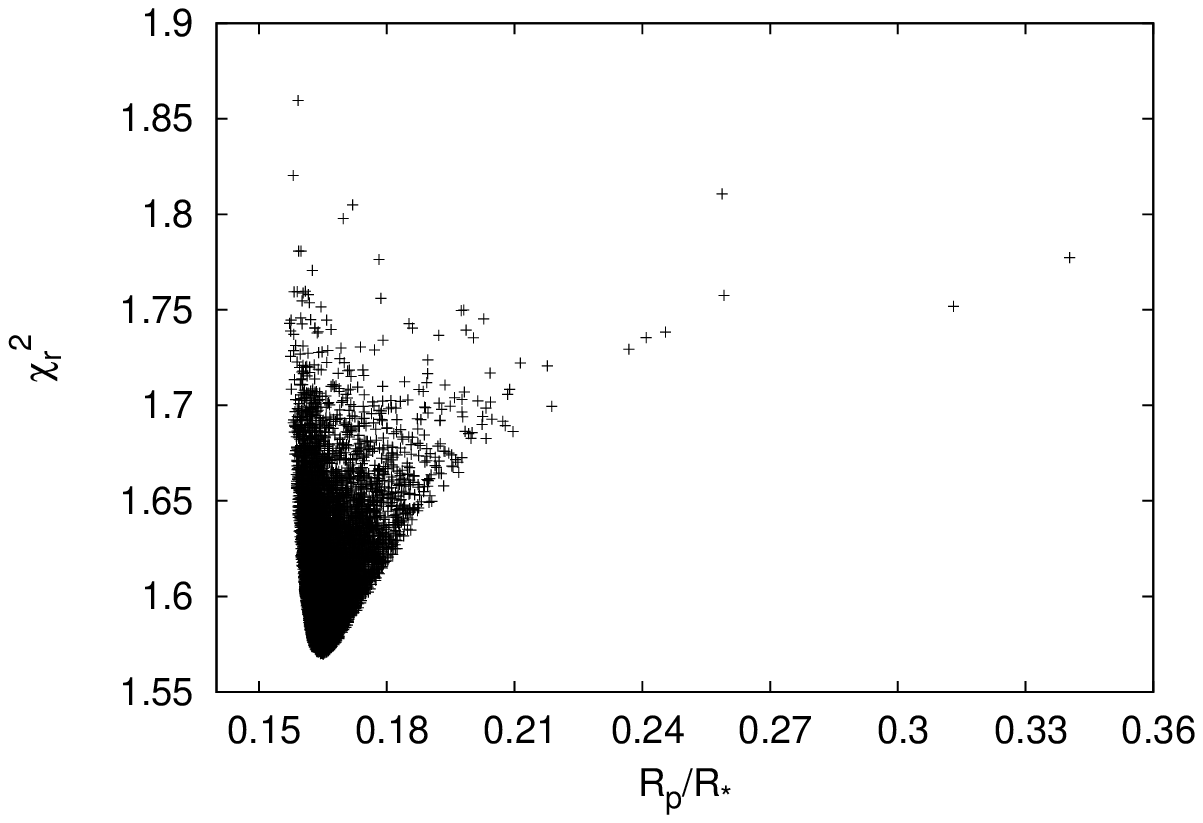}
\includegraphics[width=80mm,clip=]{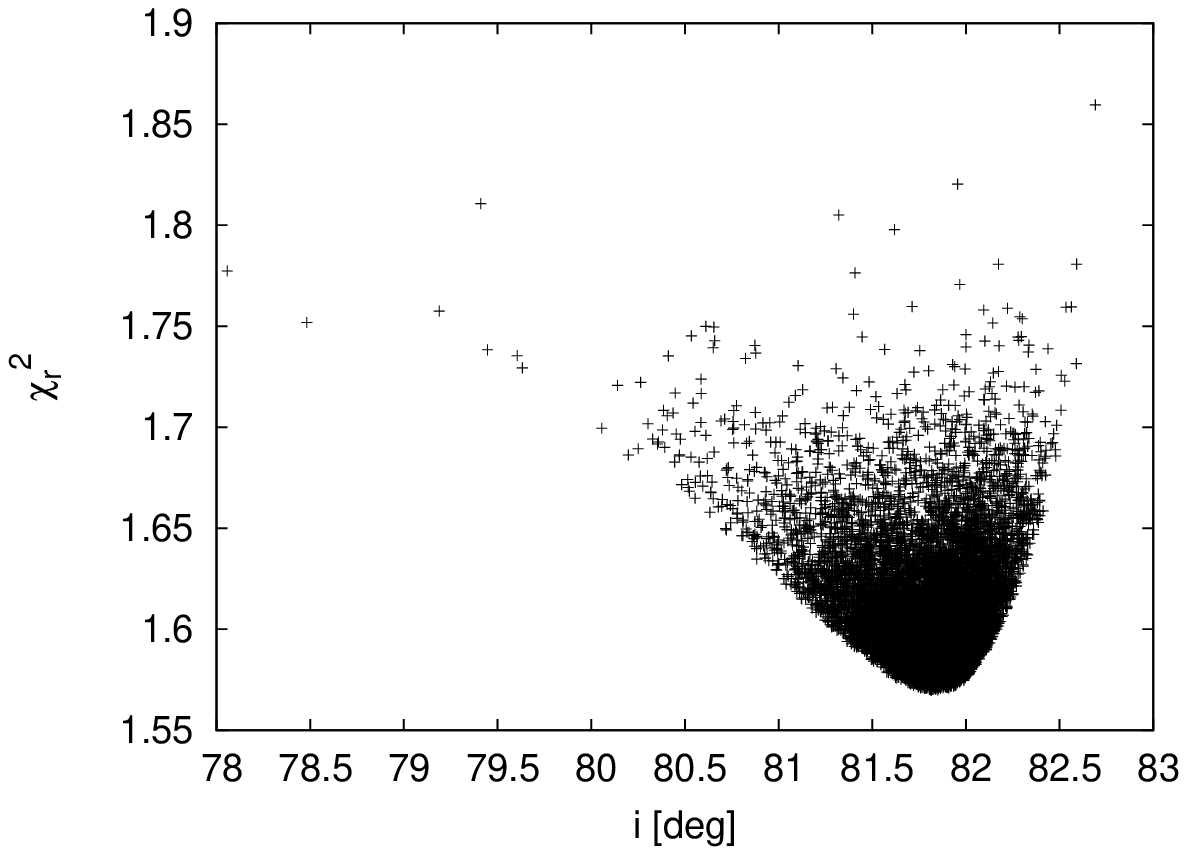}
\caption{Top: Reduced $\chi ^2_{\mathrm{r}}$ representing global minimum
solution for calculated parameter $R_{\mathrm{p}}/R_*$. Bottom: Reduced $\chi ^2_{\mathrm{r}}$
representing global minimum solution for calculated parameter $i$. \label{fig03}}
\end{figure}

To estimate the uncertainties of the calculated transit parameters, we
employed the Monte Carlo simulation method \citep{press+92}. We produced
10\,000 synthetic data sets with the same probability distribution as the residuals 
of the fit in Figure \ref{fig02}. From each synthetic data set obtained by this way we 
estimated the synthetic transit parameters. The minimum $\chi ^2_{\mathrm{min}}$ value 
corresponding to each set of Monte Carlo parameters was calculated as:
\begin{equation}
\chi ^2_{\mathrm{min}}=\sum _{i=1}^{N}
\left(\frac{m_{i}-s_{i}}{\sigma_{i}}\right)^2,
\end{equation}
where $m_{i}$ is the original best fit model value and $s_{i}$ is the 'Monte--Carlo
simulated' value. Figure \ref{fig03} shows the dependence of the parameters
$R_{\mathrm{p}}/R_{*}$ and $i$ on the reduced $\chi ^2_{\mathrm{r}}$. This quantity is defined as:
\begin{equation} 
\chi ^2_{\mathrm{r}} = \frac{\chi ^2_{\mathrm{min}}}{N-M},
\end{equation}
where $N$ is the number of data points and $M$ is the number of fitted
parameters. 
To fully understand the errors of the system parameters we constructed confidence
intervals in 2-dimensional space. Figure \ref{fig04} depicts the confidence regions for
2 parameters ($R_{\mathrm{p}}/R_{*}$ vs. $i$) as a projection of the original
4-dimensional region. The gray-colored data points stand for the $1 \sigma$,
$2\sigma$ and $3\sigma$ region with corresponding value of 
$\Delta \chi ^2 =\chi ^2_{\mathrm{min}} -\chi ^2_{\mathrm{m}} = 4.72$, $9.7$ and
$16.3$, respectively ($\chi^2_{\mathrm{m}}$ is the $\chi ^2$ of the original best fit model
value). From the shape of the dependence it could be seen that these two
parameters correlate (see Section \ref{analysis}).

Finally, we took into account the uncertainty of the stellar mass and semi-major axis according to 
the simple error propagation rule. The results from the first analysis are
shown in the Table
\ref{tab03} as SOLUTION~1. 

\subsection{SOLUTION 2 (JKTEBOP code)}
\label{jktebop}

We have modelled the light curve using the JKTEBOP\footnote{JKTEBOP
is written in FORTRAN77 and the source code is available at 
http://www.astro.keele.ac.uk/jkt/codes/jktebop.html} code
\citep{southworth11} as well. 
JKTEBOP grew out of the original EBOP program written for eclipsing binary 
systems \citep{etzel81, popper+81} via implementing the NDE \citep{nelson+72} 
model. JKTEBOP uses biaxial spheroids to model the component stars (or star and planet) 
and performs a numerical integration in concentric annuli over the surface of each 
body to obtain the flux coming from the system. This feature of the code allows
us to avoid 
small and spherical planet approximations which are used in analytic light-curve 
generators based on \citet{mandel+02}, and hence to derive planet’s oblateness. 
A model is fitted to the data by the Levenberg-Marquardt least-square procedure. 
The code converges rapidly toward a reliable solution and diminishes the correlation 
between fitted parameters \citep{southworth08}.

The main parameters of a JKTEBOP fit are the orbital inclination $i$, and 
the fractional radii of the host star and planet, $r_A$ and $r_b$. 
The fractional radii are defined as:

\begin{equation}
r_A = \frac{R_{*}}{a},~~~~~~~~r_b = \frac{R_p}{a},
\end{equation}
where $R_*$ and $R_{\mathrm{p}}$ are the stellar and planetary radii and $a$ is the 
orbital semi-major axis. Parameters $r_A$ and $r_b$ correspond to radii of spheres 
of the same volume as the biaxial spheroids. In JKTEBOP the fractional radii 
are reparametrized as their sum and ratio:

\begin{equation}
r_A + r_b,~~~~~~~~k = \frac{r_b}{r_A} = \frac{R_p}{R_*},
\end{equation} 
because these are only weakly correlated with each other \citep{southworth08}. 
The directly fitted orbital inclination, $i$, allows the transit
impact parameter $b$ = $\frac{a}{R_*}$ cos $i$ to be calculated. 
The initial values of
parameters listed above were taken from \citet{sozzetti+09}. Because of different quality of data, the
synthetic light curve was calculated only for the best light curve obtained at Calar Alto on
September 06, 2010 (the same like in SOLUTION~1). A value 
of $a$ = 0.02282$^{+0.00023}_{-0.00040}$ au \citep{sozzetti+09} was used in
subsequent calculations. The best-fit model was used as a template and
fitted to other light curves for which only the mid-transit time was allowed to vary.
The resulting mid-transit times together with $T_C$ obtained from literature 
\citep{sozzetti+09, gibson09} are analysed in details in the Section~4. 
The determined parameters obtained by JKTEBOP code are presented in 
Table~\ref{tab03} as SOLUTION~2.

The errors of derived parameters were determined in two ways for each
combination of data set and adopted LD law (we have used the same
coefficients like in the case of SOLUTION~1). Firstly, we ran 1000 
Monte Carlo (MC) simulations, a spread range of a given parameter within 
68.3\% was taken as its error estimate. Secondly, the prayer-bead method 
(e.g. \citealp{desert+11, winn+09}) was used to check whether red noise 
was present in our data. MC errors were found to be 2 -- 3 times smaller than the
values returned by the prayer bead method. This indicates that the light curve is 
affected not only by Poisson noise but also by additional correlated noise. 
Therefore, our prayer-bead error estimates were taken as our final errors.

\begin{table*}
\caption{Parameters of the extrasolar system TrES-3 from this work
(This work: SOLUTION~1 and SOLUTION~2) compared with the results from
the previous studies. $R_*/a$ is the star radius to semi-major axis ratio,
$R_p/R_*$ is the planet to star radius ratio, $i$ is the inclination of the
orbit, $T_D$ is the transit duration assuming the semi-major axis of 
$a$ = 0.02282$^{+0.00023}_{-0.00040}$ au \citep{sozzetti+09} and $P_{orb}$
is the orbital period. The orbital period in this work was fixed during
analysis. The errors of the orbital periods are in parenthesis.  
\label{tab03}}
\footnotesize 
\begin{center}
\begin{tabular}{lccccc}
\hline
\hline
Source & $R_*/a$ &  $R_p/R_*$ & $i$       & $T_D$ & $P_{orb}$ \\
       &         &            & $[\degr]$ &  [min] & [days]   \\
\hline
\hline 
\citet{odonovan+07} & 0.1650 $\pm$ 0.0027& 0.1660 $\pm$ 0.0024 & 82.15 $\pm$ 0.21 & -- & 1.30619(1)\\
\citet{sozzetti+09} & 0.1687$^{+0.0140}_{-0.0410}$ & 0.1655 $\pm$ 0.0020 & 81.85 $\pm$ 0.16 & -- & 1.30618581(51)\\
\citet{gibson09} & -- & 0.1664$^{+0.0011}_{-0.0018}$ & 81.73$^{+0.13}_{-0.04}$& 79.92$^{+1.44}_{-0.60}$ & 1.3061864(5)\\
\citet{colon+10} & -- & 0.1662$^{+0.0046}_{-0.0048}$& -- & 83.77$^{+1.15}_{-2.79}$& -- \\
\citet{southworth10} & --  & -- & 82.07 $\pm$ 0.17& -- & 1.3061864(5)\\
\citet{lee11} & -- & 0.1603 $\pm$ 0.0042 & 81.77 $\pm$ 0.14 & -- & 1.30618700(15)  \\
\citet{christiansen+11} & 0.1664 $\pm$ 0.0204 & 0.1661 $\pm$ 0.0343& 81.99 $\pm$ 0.30& 81.9 $\pm$ 1.1& 1.30618608(38)\\
\citet{southworth11} & -- & -- & 81.93 $\pm$ 0.13& -- & 1.30618700(72)\\
\citet{sada+12} & -- & -- & -- & 77.9 $\pm$ 1.9& 1.3061865(2)\\
\citet{kundurthy+13} & & & & & \\
Solution\_1          & 0.1675 $\pm$ 0.0008 & 0.1652 $\pm$ 0.0009 & 81.95 $\pm$ 0.06 & -- & 1.3062132(2) \\
\citet{kundurthy+13} & & & & & \\
Solution\_2          & 0.1698 $\pm$ 0.0014 & 0.1649 $\pm$ 0.0015 & 81.51 $\pm$ 0.14& -- & 1.3062128(2) \\
\citet{turner+13}    & 0.1721$^{+0.0054}_{-0.0052}$ & 0.1693$^{+0.0087}_{-0.0069}$ & 81.35$^{+0.63}_{-0.51}$& 81.30 $\pm$ 0.23 & 1.306 1854(1)\\
This work           &  & & & & \\
Solution\_1         & 0.1682 $\pm$ 0.0032 & 0.1644 $\pm$ 0.0047 & 81.86 $\pm$ 0.28&79.20 $\pm$ 1.38& 1.306186\\
This work           & & & & & \\
Solution\_2         & 0.1696$^{+0.0024}_{-0.0027}$& 0.1669$^{+0.0027}_{-0.0025}$& 81.76$^{+0.14}_{-0.15}$& 79.08 $\pm$ 0.72 & 1.306186\\
\noalign{\smallskip}
\hline
\hline
\end{tabular}
\end{center}
\end{table*}

\subsection{Light curve analysis results}
\label{analysis}

The resulting values of the parameters together with their uncertainties are given in
Table \ref{tab03}. For comparison, the parameters from previous studies are added
there as well. Figure \ref{fig02} shows the resulting best fit obtained by SOLUTION~1.
In order to compare both solutions, we also plotted differences from the fit
obtained by routines resulting from SOLUTION 1 and SOLUTION 2.
 
The final parameters are in good agreement with already published values. The output from SOLUTION~1,
in particular $R_{\mathrm{p}}$ and $R_*$, correspond a bit better to parameters of
\citet{sozzetti+09} and \citet{christiansen+11}. The radii determined by JKTEBOP
are also within the range of errors. Based on these two parameters we also
inferred the critical value of inclination for total transit of TrES-3b
as:

\begin{equation}
{\rm cos}~i = \frac{R_*}{a} - \frac{R_p}{a} \Rightarrow i = 81.9\degr.
\end{equation}

\noindent As listed in Table \ref{tab03}, all values of inclination including
uncertainties are in agreement with critical inclination calculated above. 
This is an evidence of grazing transit of TrES-3b.

\begin{figure}
\includegraphics[width=80mm,clip=]{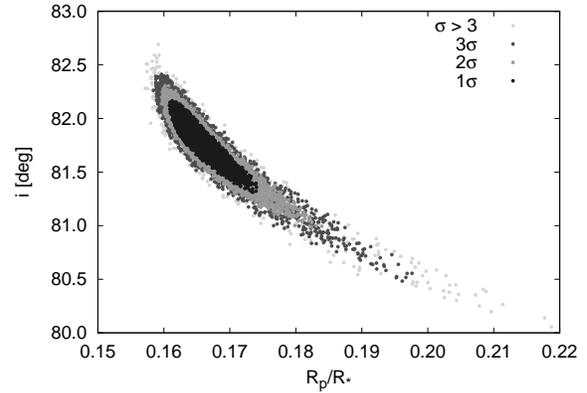}
\caption{Confidence region depicted as a projection of the 4-dimensional
region into 2-dimensional parameter space (inclination and ratio of the
planet to star radius). Regions of $1 \sigma$, $2\sigma$ and $3\sigma$ 
corresponding to $\Delta \chi = 4.72$, $9.7$ and $16.3$, respectively, 
are marked with different gray colors. }\label{fig04}
\end{figure}

We used MC simulations to produce of 10000 synthetic
data sets with the same probability distribution as the residuals
of the fit. From each synthetic data set we estimated the synthetic
transit parameters. Using the results of the simulation, the dependence
of the inclination versus the planet to star radius ratio is plotted in Figure \ref{fig04}.
The figure demonstrates that there is a correlation between the parameters and that solution 
for the system TrES-3 is not
unique and can be located in a relatively wide range (degeneracy of the
parameters). This fact is also caused by grazing transit and by subsequent sensitivity of
the solution to the determination of LD coefficients.

\begin{figure*}
\includegraphics[width=150mm,clip=]{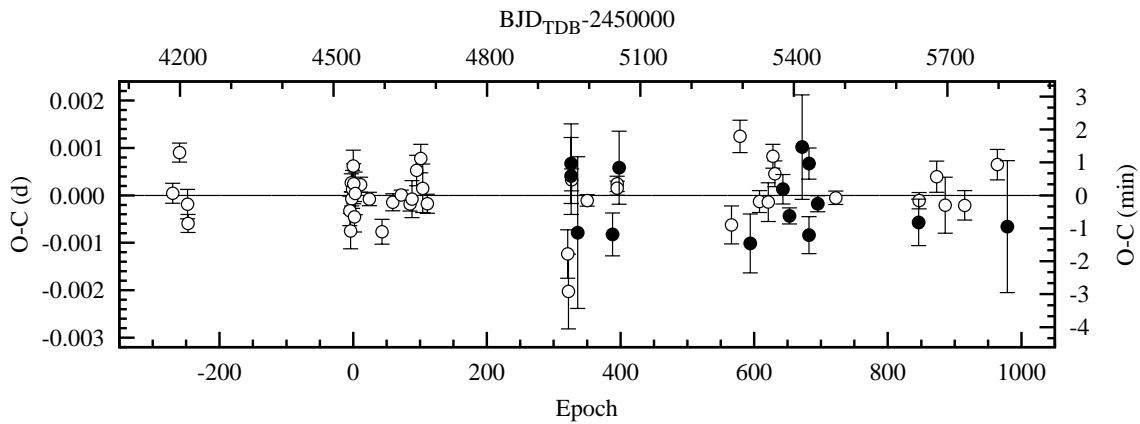}
\caption{Observation minus calculation $O-C$ diagram for transit timing of
TrES-3~b, plotted according to the new linear ephemeris. 
Open circles denote re-analysed mid-transit times from the literature.
Filled symbols mark new mid-transit times reported in this paper.}
\label{fig05}
\end{figure*}

For determination of our mid-transit times, the best-fit model was used
as a template and fitted to other light curves for which only the
mid-transit time was allowed to vary. In order to have
homogeneous analysis of TTV, we collected the data points of  
transit light curves published in previous studies (see Figure~\ref{fig05}). 
Transit light curve data of \citet{sozzetti+09} and \citet{colon+10} were
available on-line and the
remaining tabulated data were obtained from mentioned authors by e-mail. We
re-analysed all transit light curves and derived mid-transit times by the
same procedure (the code JKTEBOP) to get a homogeneous dataset for TTV
analysis (see Section \ref{ttv}). 

We also re-analysed all light curves under consideration with
$k = r_b/r_A$ as a free parameter and tried to search for any variation. 
All determinations of $k$ were consistent within error bars with a
mean value, so we did not detect any transit depth variation. 
We also saw no significant signal in periodograms.

\section{Transit Timing Variation}
\label{ttv}

Our new 14 mid-transit times and 42 redetermined literature values 
were used to refine the transit ephemeris.
The mid-transit times were transformed from JD or HJD (based on UTC) 
into BJD (based on Barycentric Dynamical Time -- TDB) using the on-line 
converter\footnote{http://astroutils.astronomy.ohio-state.edu/time/utc2bjd.html} by \citet*{Eastman10}. 
As a result of fitting a linear function of the epoch, we obtained the mid-transit time for the initial 
epoch $T_0=2454538.58144\pm0.00007$ BJD$_{\rm{TDB}}$ and the orbital period $P_{\rm{b}}=1.30618599\pm0.00000023$ d. 
The individual mid-transit errors were taken as weights. The linear fit yields reduced $\chi^2$ of $2.5$ that is 
similar to a value of 2.3 reported by \citet{gibson09}. These values, noticeable greater than 1 might suggest the 
existence of an additional planet which perturbs the orbital motion of TrES-3~b \citep{sozzetti+09,gibson09}. 

\begin{table*}
\caption{Results of transit timing. Obs. codes an observatory and instrument
according to Table ~1. $T_{0}$ denotes the
mid-transit times given as BJD (based on Barycentric Dynamical Time, TDB).
Errors of mid-transit times are in days.    
The $O-C$ values were calculated according to the new ephemeris.
\label{tab04}}
\footnotesize
\begin{center}
\begin{tabular}{lcclll}
\hline
\hline
Date &  Obs. & Epoch & $T_{0}$ $(\rm{BJD_{\rm{TDB}}})$ & $T_{0}$ error &
$O-C$ (d)   \\
\hline
2009 May 12  & P  & 326  &  2454964.39885  &  0.00084  &  $+0.00077$ \\
2009 May 12  & G1  & 326  &  2454964.39859  &  0.00081  & $+0.00051$  \\
2009 May 25  & GSH (CTK)  & 336  &  2454977.4593  &  0.0016  & $-0.0007$  \\
2009 Aug 01  & G1  & 388  &  2455045.38091  &  0.00045  &  $-0.00070$ \\
2009 Aug 14  &  VK & 398  &  2455058.44418  &  0.00077  &  $+0.00071$ \\
2010 Apr 27  & G1  & 594  &  2455314.45510  &  0.00062  &  $-0.00082$ \\
2010 Jun 30  &  G1 & 643  &  2455378.45937  &  0.00031  &  $+0.00034$ \\
2010 July 13 &  MA & 653  &  2455391.52067  &  0.00017  & $-0.00022$  \\
2010 Aug 07  &  G1 & 672  &  2455416.3397  &  0.0011  & $+0.0012$  \\   
2010 Aug 20  & MA  & 682  &  2455429.40118  &  0.00033  & $+0.00090$  \\
2010 Aug 20  &  VK & 682  &  2455429.3997  &  0.0004  & $-0.0006$  \\   
2010 Sep 06  &  CA & 695  &  2455446.38075  &  0.00017  &  $+0.00005$ \\
2011 Mar 22  &  GSH (STK) & 846  &  2455643.6145  &  0.0005  & $-0.0003$  \\
2011 Sep 12  &  CA & 979  &  2455817.3372  &  0.0014  & $-0.0003$  \\  
\noalign{\smallskip}
\hline
\hline
\end{tabular}
\end{center} 
\end{table*} 

Results for new mid-transit times are shown in Table~\ref{tab04}. The
observed minus calculated (O--C) diagram, plotted in Fig.~\ref{fig05}, 
shows no significant deviation from the linear ephemeris. 
All our data points deviated by less than $3 \sigma$. We also searched for a
periodicity that could be a sign of an additional body in the system. 
We generated a Lomb--Scargle periodogram \citep{Lomb,Scargle} for the residuals
in a frequency domain limited by the Nyquist frequency and found the highest
peak at $\sim$30 d 
and peak-to-peak amplitude of $70\pm20$ s. This period could coincide
with a stellar rotation period which 
is roughly estimated to be $\sim$28 d. Examples of such TTV signals induced
by the stellar activity are 
observed in the Kepler data (e.g. \citealp{Mazeh13}). 
However, the false alarm probability (calculated empirically by a bootstrap
resampling method with $10^5$ trials) 
of the putative signal is disqualifying with a value of 18.2\%.
In addition, the amplitude is close to the mean 1-$\sigma$ timing  error of
our observations. 
These findings allow us to conclude that a strictly periodic TTV 
signal with the amplitude greater than  $\sim$1 minute over a 4-year time
span seems to be unlikely.

Following \citet{gibson09}, we put upper constraints on a mass of a potential perturbing planet in the system with refined assumptions. 
We adopted a value of 1 min for the maximal amplitude of the possible TTV signal. We also increased the sampling resolution to probe 
resonances other than inner and outer 1:2 commensurabilities studied in detail by \citet{gibson09}. We simplified the three-body problem 
by assuming that the planetary system is coplanar and initial orbits of both planets are circular. The masses of both the star and the 
transiting planet, as well as its semi-major axis were taken from \citet{sozzetti+09}. The orbital period of the hypothetical perturber 
was varied in a range between 0.2 and 5 orbital periods of TrES-3~b. We produced 2800 synthetic O--C diagrams based on calculations done 
with the \textsc{Mercury} code \citep{chambers99}. We applied the Bulirsch--Stoer algorithm to integrate the equations of motion. 

\begin{figure}
  \begin{center}
   \includegraphics[width=85mm,clip=4]{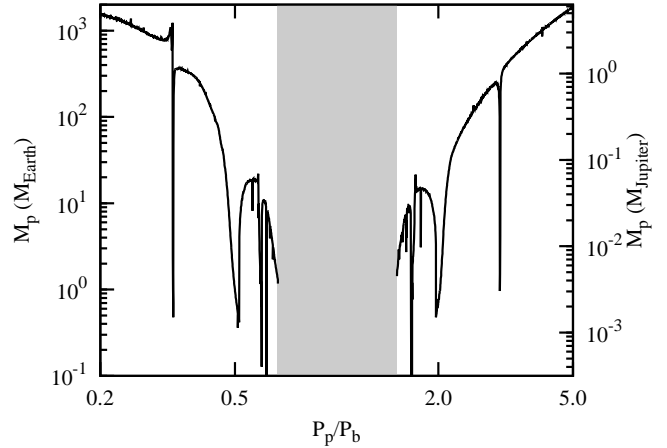}
  \end{center}
\caption{The upper-mass limit of a hypothetical additional planet that could
perturb the orbital motion of TrES-3b as a function of
ratio of orbital periods of transiting planet, $P_{\rm{b}}$, and the
hypothetical  perturber, $P_{\rm{p}}$.
Orbits located in a grey area were found to be unstable due to close
encounters of both planets.}
\label{fig06}
\end{figure}

The most important feature of the Bulirsch-Stoer algorithm 
for N-body simulations is that it is capable of keeping an upper 
limit on the local errors introduced due to taking finite time-steps 
by adaptively reducing the step size when interactions between the particles
increase in strength.
Calculations covered 1500 days, i.e. a total time span of available transit observations. The results of simulations are presented in 
Fig.~\ref{fig06}. Our analysis allows us to exclude an additional Earth-mass planet close to inner 
1:3, 1:2, and 3:5 and outer 5:3, 2:1, and 3:1 mean-motion resonances (MMRs).

\section{Long-term stability of the system}
In this section, we investigated the long-term gravitational influence of TrES-3b planet
on a potential second planet in the system. Thus, we performed numerical simulation 
for studying the stability of orbits and checking their chaotic behavior using 
the method of maximum eccentricity (e.g. \citealp{dvorak03}).

We used long-term integration of small-mass (Earth-mass) planet orbits 
for inspecting the stability regions in the TrES-3b system. 
The time span of the integration was around 140000 revolutions 
of the planet around the star. For the integration 
of this system we again applied an efficient variable-time-step 
algorithm (Bulirsch-Stoer integration method). The parameter $\epsilon$ 
which controls the accuracy of the integration was set to $10^{-8}$, 
in this case.
\begin{figure*}
 \begin{center}
  \centerline{\includegraphics[width=95mm]{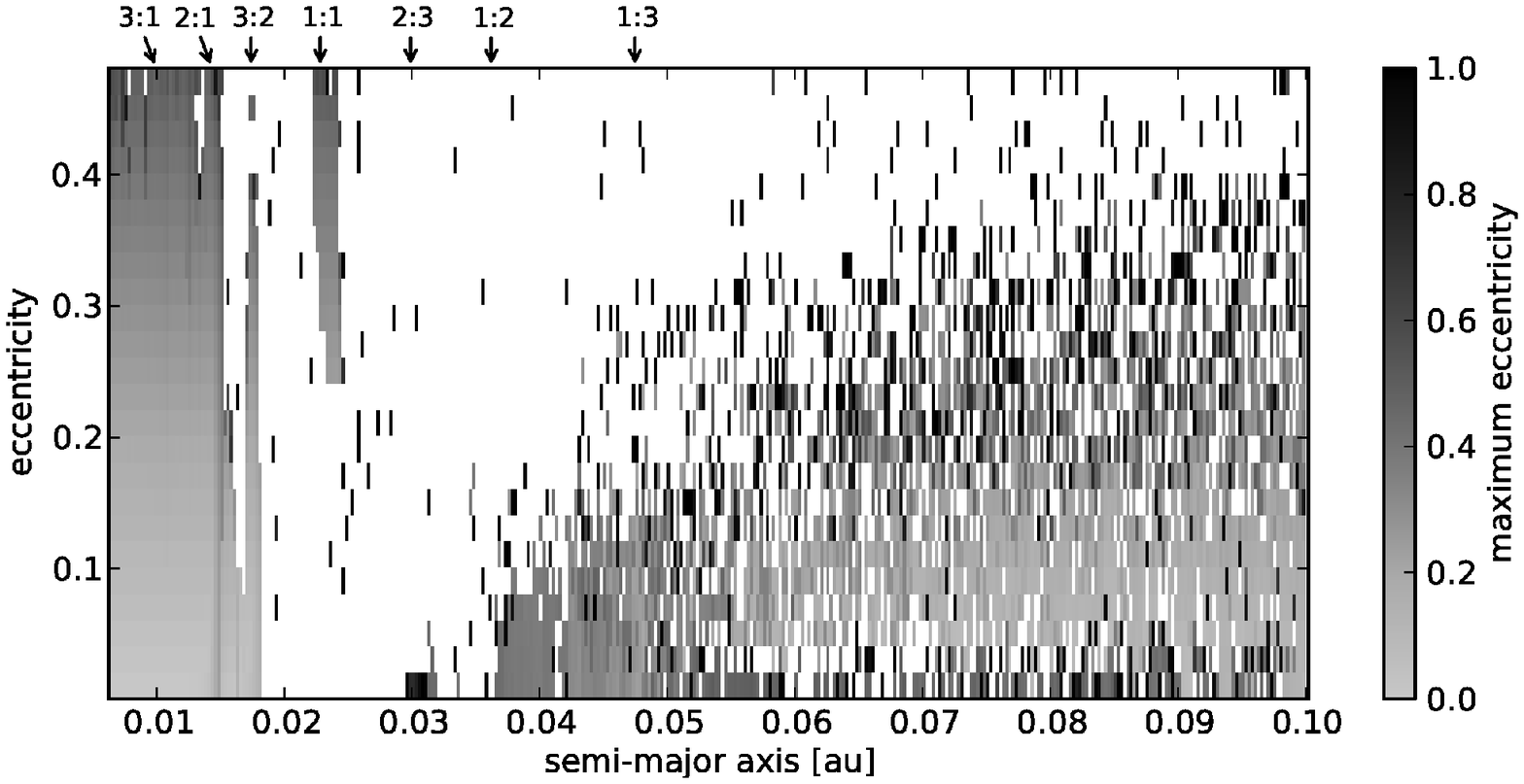}
  \includegraphics[width=95mm]{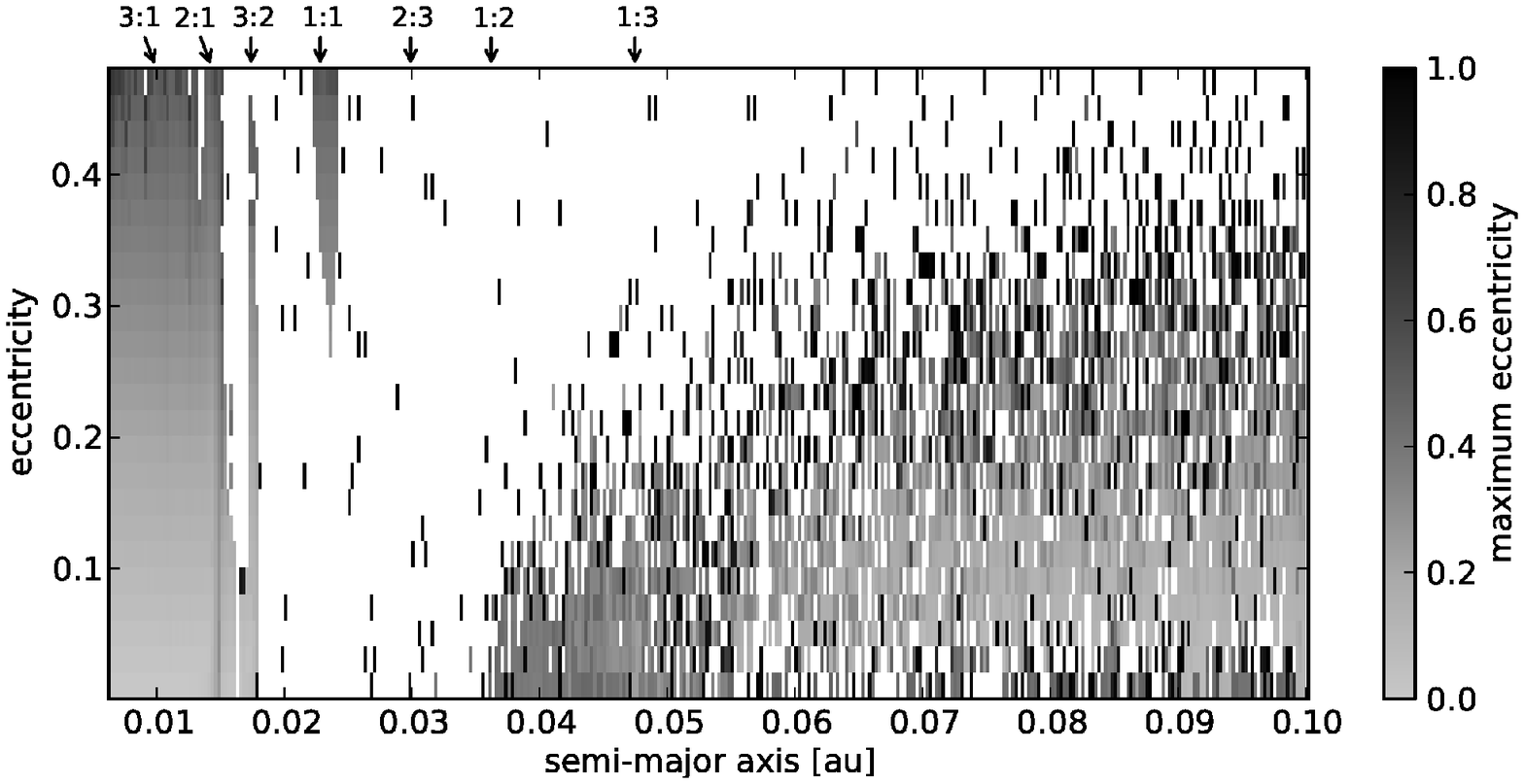}}
  \centerline{\includegraphics[width=95mm]{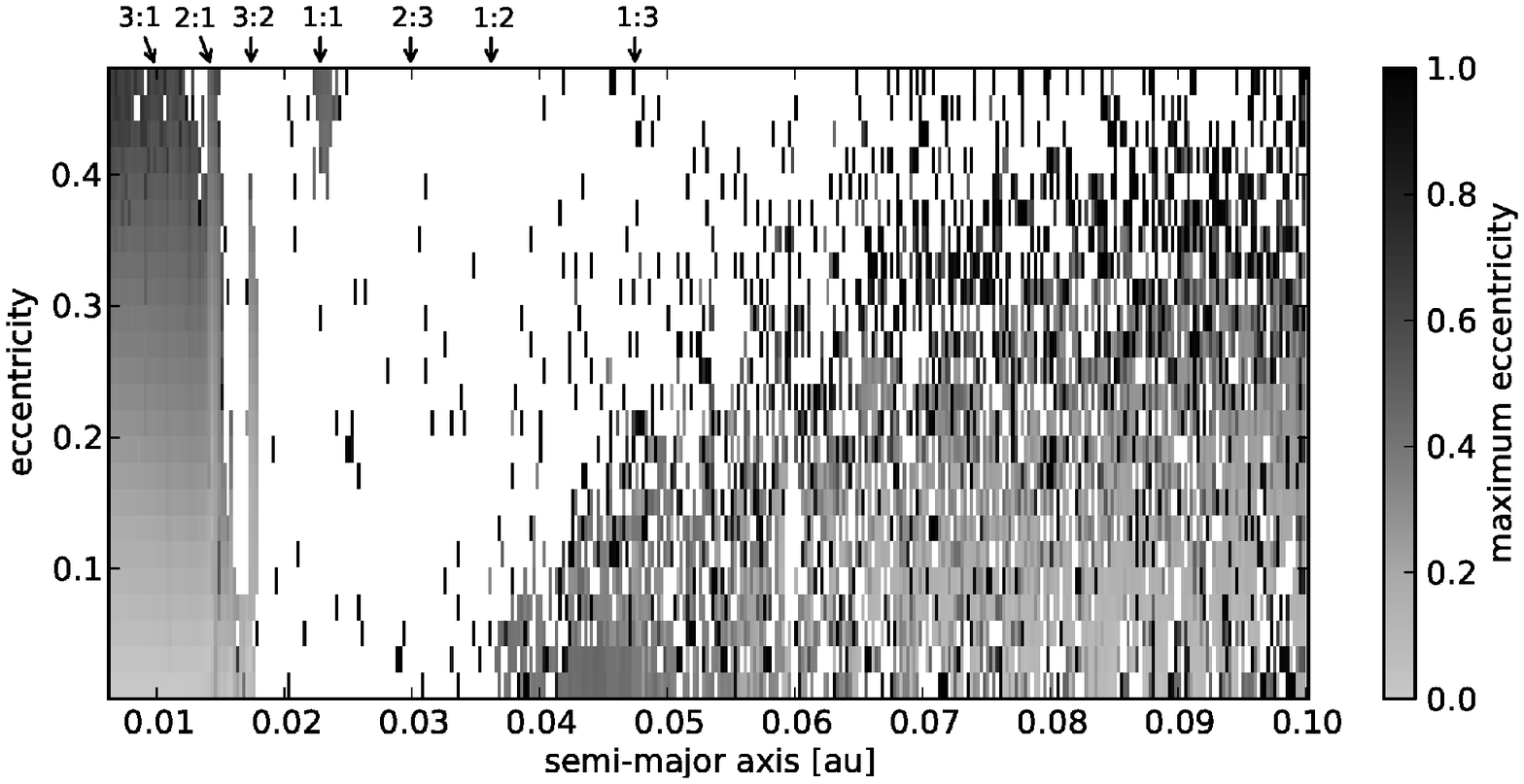}
  \includegraphics[width=95mm]{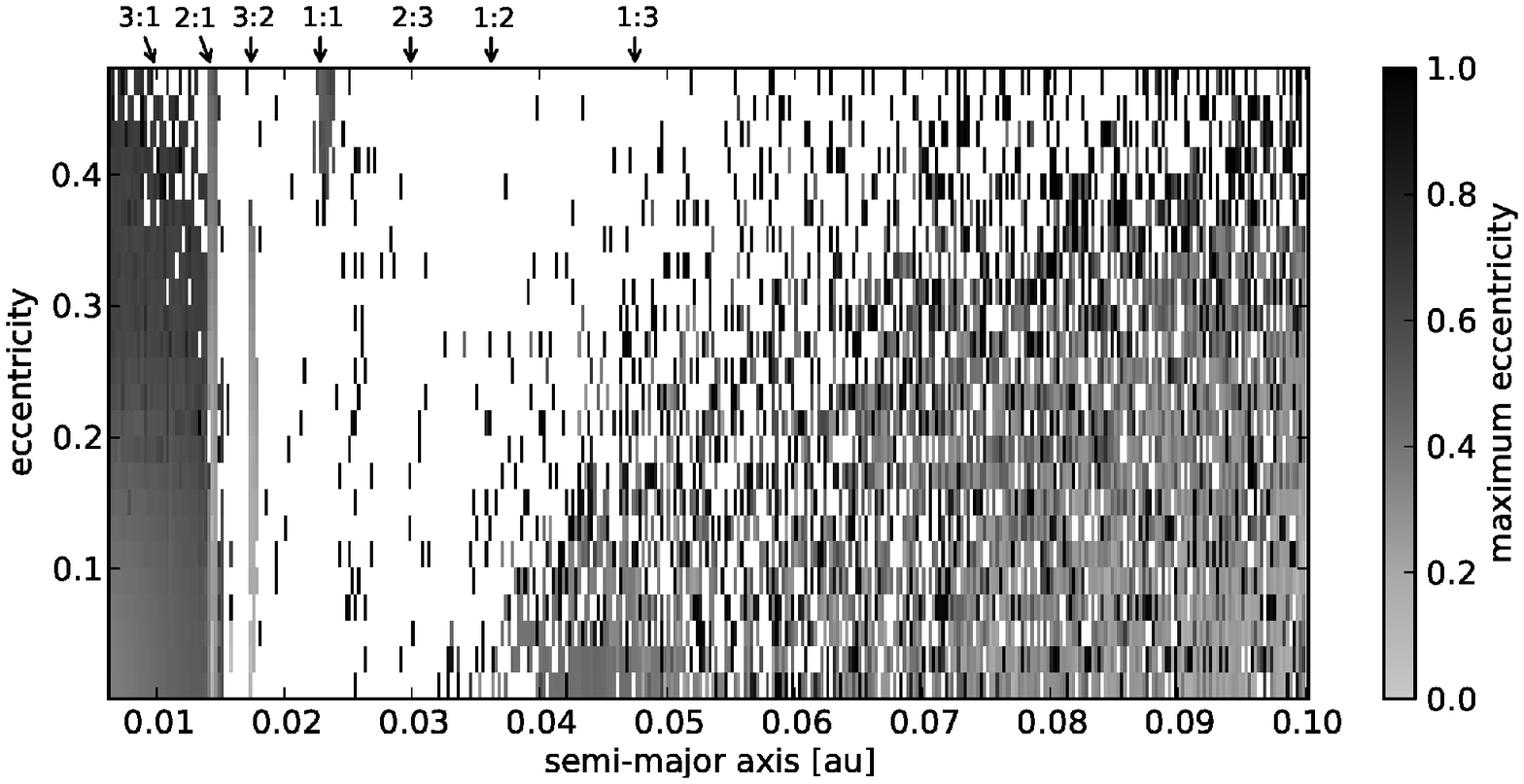}}
  \caption{Stability plot in the $a-e$ plane showing the maximum
eccentricity. From top left to bottom right: $i=5\degr$, $i=20\degr$,
$i=35\degr$ and
$i=50\degr$. The mean-motion resonances with TrES-3b planet are also marked.
The minimal value of semi-major axis is 0.00625\,au.}
  \label{fig_ae}
 \end{center}   
\end{figure*}
\begin{figure*}
 \begin{center}
  \centerline{\includegraphics[width=95mm]{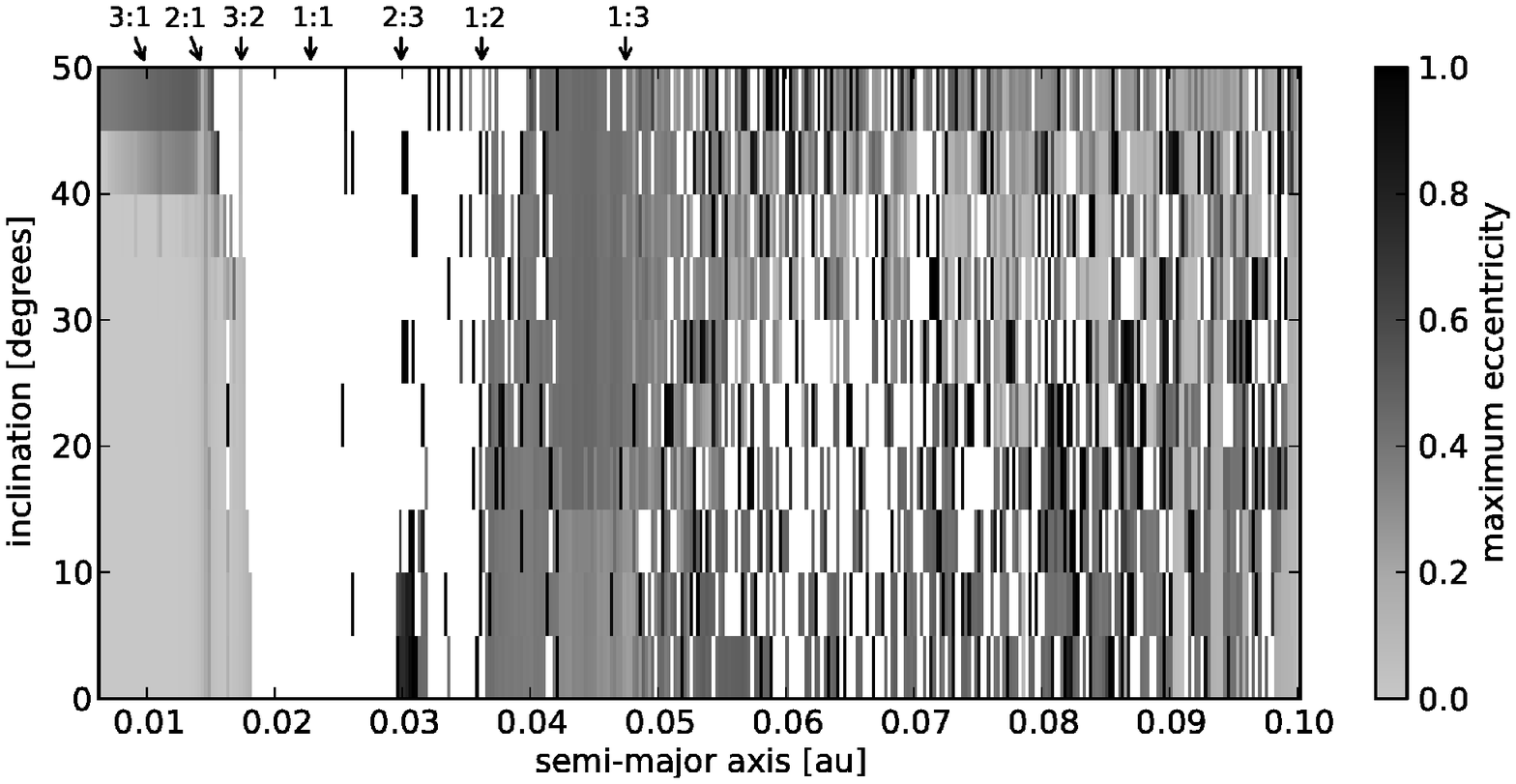}
  \includegraphics[width=95mm]{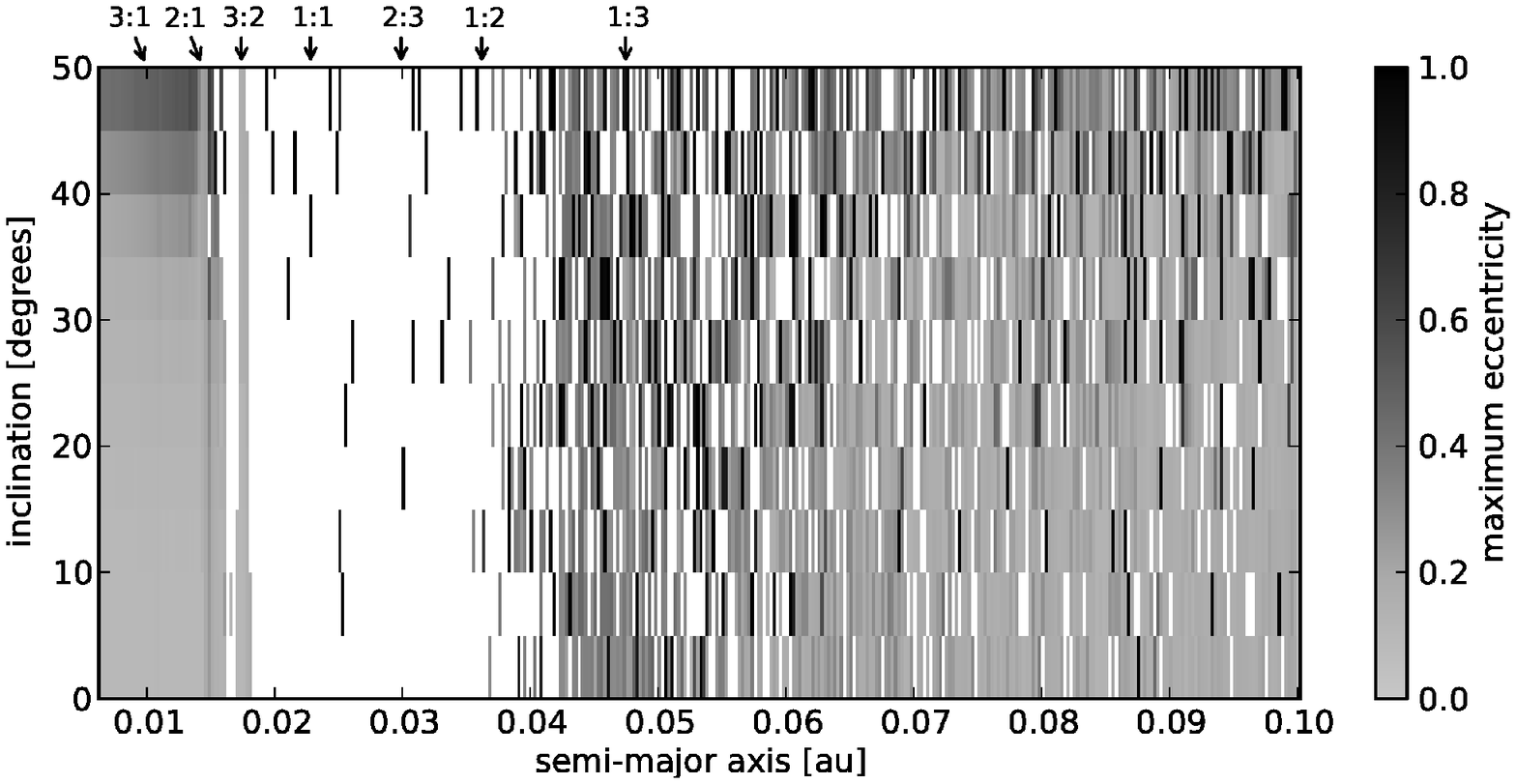}}
  \centerline{\includegraphics[width=95mm]{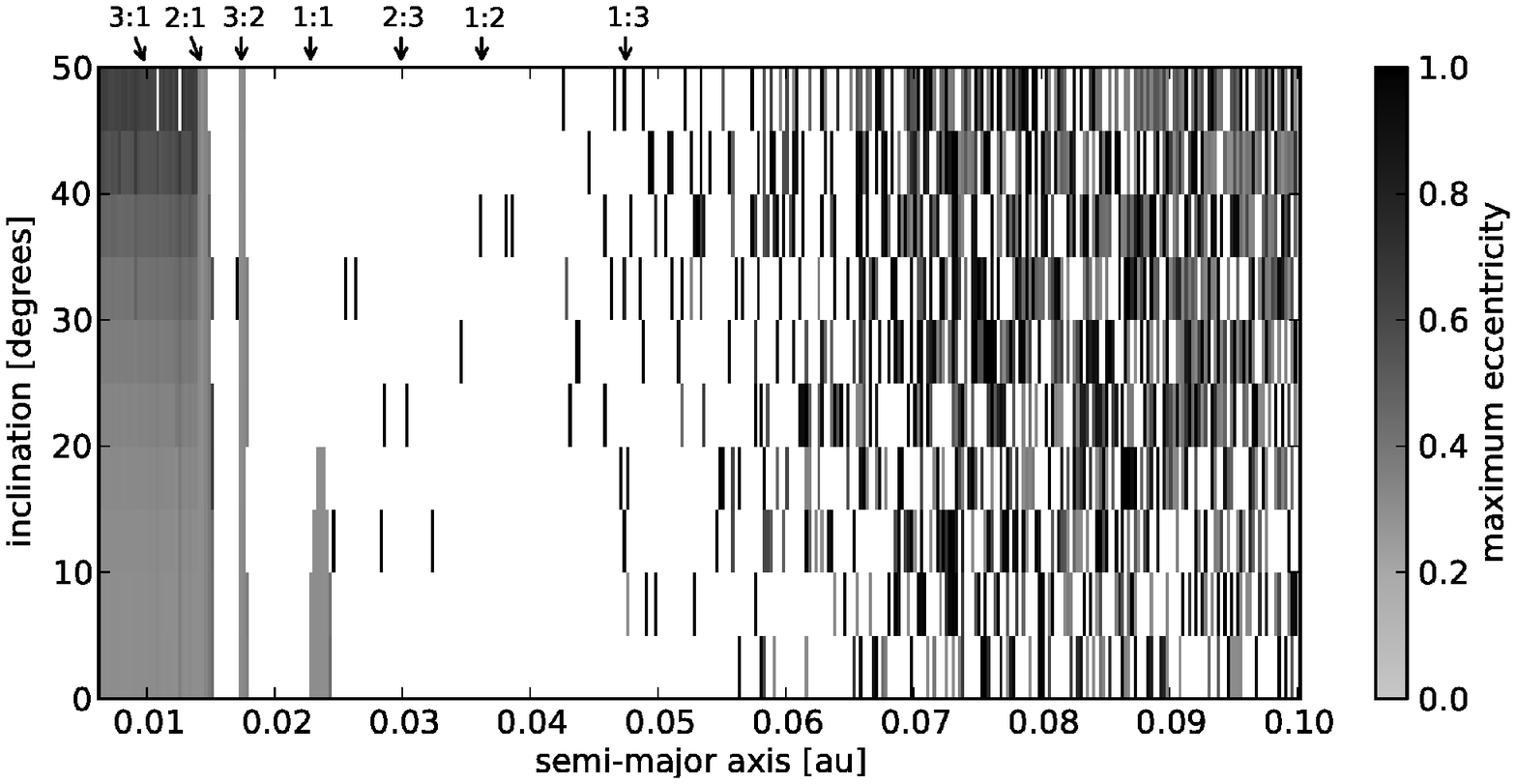}
  \includegraphics[width=95mm]{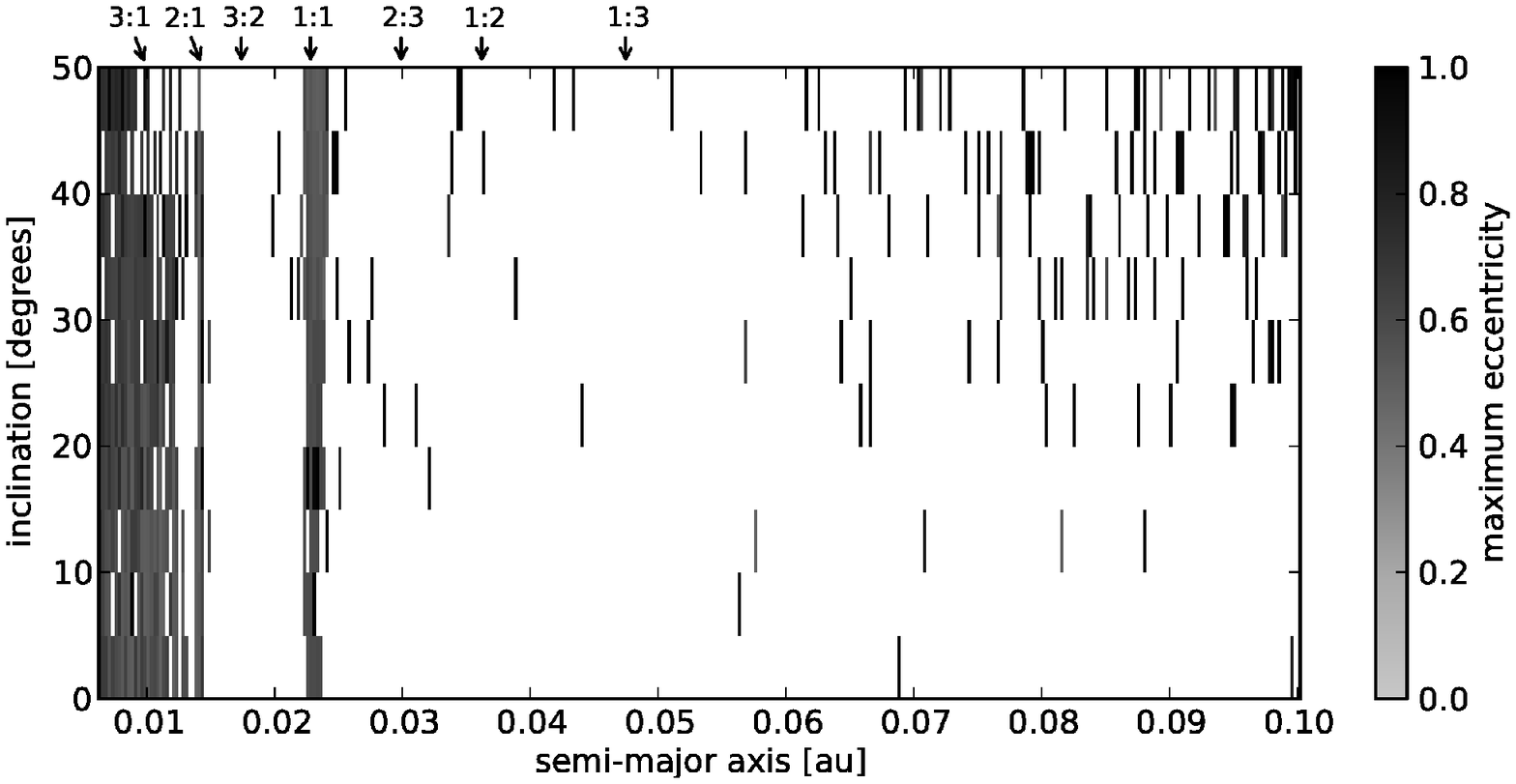}}
  \caption{Stability plot in the $a-i$ plane showing the maximum
  eccentricity.
From top left to bottom right: $e=0.001$, $e=0.1$, $e=0.3$ and $e=0.5$.
  The mean-motion resonances with TrES-3b planet are also marked.
  The minimal value of semi-major axis is 0.00625\,au.
}
  \label{fig_ai}
 \end{center}   
\end{figure*}
We have generated $10^5$ mass-less particles for representing small planets 
in this system. We assumed that semi-major axis ranges from 0.00625$\,$au
to 0.1$\,$au. The inner border of the generated system of small planets is around 1.5 times
star radius. This value of inner border is approximately at the location of Roche limit 
for a Earth-like planet and includes also the region in the vicinity of 3:1 MMR 
(0.01\,au for this case) in our analysis.
The step-size in semi-major axis was $\Delta\,a$ = 0.00025$\,$au, in eccentricity 
$\Delta\,e$ = 0.02 and $\Delta\,i$ = 5$\degr$ in inclination. 
The upper limit of the grid in eccentricity was 0.5 and 50$\degr$ in inclination.
We integrated the orbits of the small planets for the time-span of 500 years 
(about 140 000 revolutions of TrES-3b around the parent star). 
We obtained orbital evolution of each small planet in the system and also
were able to study the stability regions at the end of the integration
using the method of maximum eccentricity, mentioned above. 
The maximum eccentricity that the potential perturber of TrES-3b
reached over the time of the integration is plotted on the $a-e$ ($a-i$) stability
map in the Fig.\,\ref{fig_ae} (Fig.\,\ref{fig_ai}).

Figure \ref{fig_ae} shows the stability maps in $a-e$ plane 
for selected values of inclinations: $i=5\degr$, $i=20\degr$, $i=35\degr$ 
and $i=50\degr$ (from top left to bottom right). Also the MMRs 
with TrES-3b planet are marked in these plots. We can see a stable 
region inside the 2:1 MMR for each value of inclination.
Outside the 2:1 MMR the gravitational influence of the TrES-3b planet 
is very strong and leads to depletion of these regions. Only the MMRs 
2:1, 3:2 and 1:1 are moderately populated but the population decreases 
with the increase of the inclination. For completeness, we note 
that 5:3 ($a \approx$ 0.016\,au) and 3:5 ($a \approx$ 0.032\,au) MMRs are 
depleted and thus not stable for additional Earth-mass planet in their 
vicinity.
In Fig.\,\ref{fig_ai} we present the stability maps in $a-i$ plane 
for several values of eccentricities: $e=0.001$, $e=0.1$, $e=0.3$ and $e=0.5$. 
One can see the stable region inside the 2:1 MMR with TrES-3b planet, which
we found stable also in the $a-e$ plane. Considering the region beyond the
1:3 MMR, the depletion of the planet population is not so strong than
in the region between 2:1 and 1:3 MMR. This feature is in a good agreement 
with the weaker gravitational influence of the TrES-3b planet.

Based on our results, we showed that the region inside the TrES-3b planet orbit 
(especially inside the 2:1 up to 3:1 MMR) can be stable on longer timescales (hundreds of years 
or hundred thousands of revolutions of TrES-3b around the star) from the dynamical 
point of view. The region from 0.015$\,$au (near 2:1 MMR) to 0.05$\,$au (near 1:3 MMR) 
was found unstable apart from moderately populated MMRs located in this area 
(see figures Fig.\,\ref{fig_ae} and Fig.\,\ref{fig_ai}). The relatively small increase 
of population in the mentioned MMRs depends on the initial values of semi-major axis, 
eccentricity and inclination. The region beyond the 0.05$\,$au was found to have a chaotic 
behavior and the depletion of the planet population increases with increasing values 
of initial eccentricity as well as inclination.

\section{Discussion and Conclusion}

Based on the transit light curves obtained at several
observatories between May 2009 and September 2011 we redetermined orbital 
parameters and the radius of the transiting planet TrES-3b. The best light curve 
(obtained at Calar Alto, Sep 2010) was used for light curve analysis, and the data from 
the other observatories were used for $T_{C}$ determination and TTV investigation. 
We used two independent solutions for parameters determination and finally, 
we concluded that our values are consistent with previous results 
of \citet{sozzetti+09} and \citet{christiansen+11}. 

The aim of this present paper was also to discuss possible presence 
of a second planet in the extrasolar system TrES-3. For this purpose we used 
our new 14 mid-transit times and the individual determinations from
\citet{sozzetti+09} and \citet{gibson09}. The resulting $O-C$ diagram showed 
no significant deviation of data points from the linear ephemeris.   
In addition, we tried to search for a periodicity that could be caused by
an additional body in the system. We can conclude that a strictly
periodic TTV signal with the amplitude greater than 1 minute over a 4-year time span 
seems to be unlikely. This result, together with refined assumptions of
\citet{gibson09} allow us to put upper constraints on the mass of a potential
perturbing planet. The additional Earth-mass planet 
close to inner 3:1, 2:1, and 5:3 and outer 3:5, 1:2, and 1:3 MMRs 
can be excluded.

Finally, we used the long-term integration of the theoretical set of massless
particles generated in TrES-3 system for studying the dynamical stability 
of potential second planet in the system (influenced by TrES-3b gravitation). 
From our analysis we found that the region inside the TrES-3b planet orbit
(especially inside the 2:1 MMR) up to 3:1 MMR can be stable on longer timescales 
(hundreds of years or hundred thousands of revolutions of TrES-3b around the star) 
from the dynamical point of view. The region from 0.015$\,$au (near 2:1 MMR) to 
0.05$\,$au (near 1:3 MMR) was found to be unstable apart from moderately 
populated MMRs located in this area. The relatively small increase of
population in these MMRs depends on the initial values of semi-major axis, eccentricity
and inclination. The region beyond 0.05$\,$au was found to have a chaotic behavior and
depletion of the planet population increases with increasing values of initial eccentricity 
as well as inclination.

\section*{Acknowledgements}

This work has been supported by the VEGA grants No. 2/0078/10, 2/0094/11, 
2/0011/10.  MV, MJ, JB, TP and \v{S}P would like to thank also the project 
APVV-0158-11. TK thanks the Student Project Grant at MU MUNI/A/0968/2009 and the
National scholarship programme of the Slovak Republic. GM acknowledges Iuventus Plus 
grants IP2010 023070 and IP2011 031971. SR thanks the German National Science
Foundation (DFG) for support in project NE~515~/~33-1.

\label{lastpage}
\end{document}